%% file: conference_101719.tex
\documentclass[conference]{IEEEtran}
\IEEEoverridecommandlockouts
\usepackage{cite}
\usepackage{amsmath,amssymb,amsfonts}
\usepackage{algorithmic}
\usepackage{graphicx}
\usepackage{textcomp}
\usepackage{xcolor}
\usepackage{psfrag}
\usepackage{pstool}
\usepackage{textcomp}
\usepackage[absolute]{textpos}
\def\BibTeX{{\rm B\kern-.05em{\sc i\kern-.025em b}\kern-.08em
    T\kern-.1667em\lower.7ex\hbox{E}\kern-.125emX}}
\newcommand{\copyrightstatement}{
    \begin{textblock}{15}(0.3,0.2)    
         \noindent
         \centering
         \textblockcolour{white}
         \footnotesize
         \copyright 2022 IEEE. Personal use of this material is permitted. Permission from IEEE must be obtained for all other uses, in any current or future media, including reprinting/republishing this material for advertising or promotional purposes, creating new collective works, for resale or redistribution to servers or lists, or reuse of any copyrighted component of this work in other works.
    \end{textblock}
}
\begin{document}

\copyrightstatement

\title{A Bit Stream Feature-Based Energy Estimator for HEVC Software Encoding\\
\thanks{This work was funded by the Deutsche Forschungsgemeinschaft (DFG, German Research Foundation) – Project-ID 447638564.}
}

\author{\IEEEauthorblockN{{ Geetha Ramasubbu, \ Andr\'e Kaup, \ Christian Herglotz }}
\IEEEauthorblockA{\textit{Multimedia Communications and Signal Processing} \\
\textit{Friedrich-Alexander University Erlangen-Nürnberg (FAU)}\\
Erlangen, Germany\\
\{geetha.ramasubbu, andre.kaup, christian.herglotz\} @fau.de.}
}
\maketitle

\begin{abstract}
The total energy consumption of today's video coding systems is globally significant and emphasizes the need for sustainable video coder applications. To develop such sustainable video coders, the knowledge of the energy consumption of state-of-the-art video coders is necessary. For that purpose, we need a dedicated setup that measures the energy of the encoding and decoding system. However, such measurements are costly and laborious. To this end, this paper presents an energy estimator that uses a subset of bit stream features to accurately estimate the energy consumption of the HEVC software encoding process. The proposed model reaches a mean estimation error of 4.88\% when averaged over presets of the x265 encoder implementation. The results from this work help to identify properties of encoding energy-saving bit streams and, in turn, are useful for developing new energy-efficient video coding algorithms.
\end{abstract}


\section{Introduction}
\label{sec:intro}
The dawn of portable devices, unrestricted Internet, and various video-focused social networking services has increased Internet video traffic drastically \cite{ciscoWhitepaper1823}. Such a tremendous increase in video data traffic leads to huge storage costs and increased server-side energy consumption for video content creation. The total energy consumption of current coding systems is globally significant as online video contributes to 1\% of total global $\mathrm{CO_2}$ emissions \cite{shiftProject19}. Furthermore, mobile devices pose limitations in terms of battery power. Notably, video encoding has considerable power requirements, which poses a problem to battery-powered devices \cite{Sharrab13}, where the battery drains fast due to increased power requirements. 

Furthermore, the compression methods used for encoding have evolved considerably in recent years and provide a greater number of compression methods. However, their processing complexity has also greatly increased \cite{VVCComplexity}, leading to a significant increase in the energy demand. Therefore, we need powerful and energy-efficient video codecs and energy-aware video-based services in modern video communication applications. 

A limitation in searching for energy-efficient algorithms is that energy measurements are complex and laborious. Hence, we need simple energy estimators to overcome the drawback of complex measurements. An extensive body of literature on decoding energy modeling, such as in \cite{Li12} and \cite{Raoufi13} exists. The decoding energy estimation model based on various characteristics, such as decoding time \cite{Herglotz15b}, the sequence characteristics such as the number of frames, resolution, and Quantization Parameter (QP) \cite{Herglotz15a}, and the number of instructions  \cite{Herglotz17a} were introduced. 
Similarly, the decoding energy estimation model based on encoded bit stream features for various decoder implementations estimates the true decoding energy with less than 8\% mean estimation error \cite{Herglotz16b, Herglotz18Journal}. 

Few recent works have explicitly addressed the encoder's processing energy. For example, \cite{Rodriguez15} established a relationship between the quantization, spatial information, and coding energy for the intra-only HEVC encoder but did not consider the presets which quantify the encoding complexity and compression performance. Furthermore, Mercat et al. measured the energy of a software encoder for many different sequences and encoding configurations \cite{Mercat17} but did not present any encoding energy estimation model. In the end, \cite{Ramasubbu22} introduces a lightweight encoding time based encoding energy estimator, which uses the \textit{ultrafast} presets' encoding time to estimate the encoding energy demand of the other presets and achieves a mean estimation error of 11.35\% when averaged over all the presets. 

To this end, this work explores the feasibility of estimating the HEVC software encoders' energy demand using the bit stream features used for decoding energy estimation in \cite{Herglotz16b, Herglotz18Journal}. Last but not least, we introduce a subset of bit stream features that could be used to estimate the encoding energy with higher accuracy than the other encoding energy estimators in the literature. Some applications of this work are joint modeling of the energy in the encoder-decoder chain using bit stream features and identifying the features and properties that significantly contribute to energy consumption. 

The rest of this paper is structured as follows: Section \ref{sec:LR} presents existing encoding energy estimation models. Section \ref{sec:features} introduces the bit stream features and their categories and the bit stream feature count. Further, Section \ref{sec:propModel} introduces the proposed model. Then Section \ref{sec:eval} introduces the energy measurement setup, sequences and encoding configurations used, and the evaluation method and then discusses the results. Lastly, Section \ref{sec:concl} concludes this work.



\section{Literature review of encoding energy estimation models}
\label{sec:LR}
This section presents existing models for encoding energy estimation. First, \cite{Rodriguez15} introduces a quantization parameter (QP) based model for intra-coding, which is a static model to estimate encoding time based on the QP and then uses this model to estimate the encoding energy demand. A generalized version of such a model of video sequence classes \cite{CTC} reads as follows: 
\vspace{-0.05cm}
\begin{equation}\label{timemodel}t_{\mathrm{enc}}=  \kappa \cdot QP^{3}- \lambda \cdot QP^{2} - \mu \cdot QP+ T_{0},\end{equation}
where $\kappa$, $\lambda$, $\mu$ are the coefficients of the model, QP corresponds to the quantization parameter, $T_{0}$ is an offset.
\begin{equation}\label{QPModel}E_{\mathrm{enc}}=t_{\mathrm{enc}} \cdot P_{\mathrm{avg}},\end{equation}
where $t_{\mathrm{enc}}$ is the estimated processing time from \eqref{timemodel} and $P_{\mathrm{avg}}$ is the mean processing power of encoder. This model is limited to for all-intra encoding and reported relatively accurate estimations for low and high values of QP, i.e., 0–15 and 40–51, but for the mid-range QP values, the errors increase above 15\% \cite{Rodriguez15}. Notably, the QPs at which the model reports a low error are seldom used. In addition, this energy model estimates the energy only for classes B, C, and D of JVET common test conditions \cite{CTC}. 
Furthermore, \cite{Ramasubbu22} introduces two models based on the encoder processing time. The first model estimates the encoder energy demand by exploiting the linear relation between encoding time and encoding energy, such that the energy can be estimated by
\begin{equation}\label{timeModel} \hat E_{\mathrm {enc}} = E_{0} + P \cdot t_{\mathrm {enc}}, \end{equation}
where $t_{\mathrm{enc}}$ is the sequence-dependent encoder processing time. The parameter $P$ (slope) can be interpreted as the linear factor representing the mean encoding power and $E_{0}$ as a constant offset. The estimates can only be obtained when the encoding process is executed once on the target device because the encoder processing time needs to be measured. Hence, this model is adapted to  perform a-priori energy estimation, i.e., energy estimation without the need to execute the encoder. Thereby, \cite{Ramasubbu22} adapts the model \eqref{timeModel} to estimate the encoding time for each x265 preset using the encoding time of the lightweight encoding process  i.e., $\textit{ultrafast}$ preset, which is less costly to obtain than \eqref{timeModel},  
\begin{equation} \label{model2}\hat{E}_{\mathrm{enc}} = E_{0}+ P \cdot t_{\mathrm{enc,uf}}, \end{equation}
where $E_{0}$ is the offset energy,  $P$ indicates the slope, and $t_{\mathrm{enc,uf}}$ denotes the encoding time consumed for a given sequence to be encoded at a given CRF using the \textit{ultrafast} preset. However, these models do not reach a mean estimation errors of less than 15\%.

\section{bit stream Features}\label{sec:features}
A bit stream is a sequence of bits representing the coded video sequences and associated data and describes the execution of a standardized decoding process \cite{HEVCRec}. Its syntax is specified in syntax structures, representing the logical entity of the information coded into the bit stream \cite{HEVC_CodingToolsBook}. The decoding process takes the syntax elements of the bit stream as input and reconstructs the video sequence according to the semantics of the coded syntax elements \cite{HEVC_CodingToolsBook}. 

A bit stream feature can be associated with a subprocess in the decoding process\cite{HEVCRec}. While decoding a single bit stream, the decoder executes these sub-processes multiple times. Each sub-process consumes specific processing energy upon each execution \cite{Herglotz16b}. Furthermore, \cite{Herglotz16b} defines a  set of features for typical sub-processes, meaning that the complete decoding process can be split into sub-processes associated with the bit stream features. Another property of these features is that they can be linked to syntax elements defined in the standard \cite{HEVCRec}. Eventually, the occurrence and value of these syntax elements and variables can be used to determine how often the bit stream features occur and hence, how often the corresponding sub-processes are executed.   By analyzing the occurrence and value of HEVC syntax elements or variables, it is possible to determine how often a bit stream feature occurs. Hence, for a given HEVC-coded bit stream, it is counted how often a sub-process is executed. Therefore, by counting the number of executed sub-processes and determining the mean processing energy, we can estimate the total decoding energy related to that sub-process \cite{Herglotz16b}. 



The set of bit stream features as used for decoding energy estimation \cite{Herglotz16b} is divided into five categories. The general features $E_0$ and the number of frames (Islice, PBslice) comprise sub-processes associated with global coding processes such as initialization of the decoding process and the slices. The former corresponds to the offset energy required for starting and ending the decoding process. The latter, the number of frames, represents the processing energy used to initialize a frame. The intraframe coding features relate to intra-prediction sub-processes of a block and corresponding flags. The interframe coding features describe the inter-prediction subprocess of a block, parsing, and fractional pel filtering, where the number of pels that need to be predicted and filtered is counted, which is counted twice in bi-prediction. The residual coding features represent the coefficient parsing of residual coefficients and transformation block size. Finally, the in-loop filtering features describe deblocking and sample-adaptive offset (SAO) filters in the in-loop filtering processes.
\vspace{-0.07cm}
\section{Proposed Encoding Energy Model}\label{sec:propModel}
In \cite{Herglotz16b}, for any given HEVC-coded bit stream, it is counted how often a sub-process is executed. The product of the number of sub-process executions (feature numbers) with its corresponding energy demand yields the required decoding energy related to this sub-process. In the end, the sum of all such energies for all bit stream features yields the estimated energy demand of the decoding process \cite{Herglotz16b}: 
\begin{equation}
  \hat E_\mathrm{dec} = \sum_{\forall i} n_i \cdot e_i, 
  \label{eq:E_gen}
\end{equation}
where the index $i$ denotes the bit stream feature index, $n_i$ number of occurrences of the bit stream feature (feature number), 
and $e_i$ the corresponding feature-specific decoding energy \cite{Herglotz16b}. Similar to the bit stream feature based model proposed in \cite{Herglotz16b}, we here propose a bit stream feature based model for the HEVC software encoding process. For an HEVC-coded bit stream,  the encoding energy estimation with the help of the feature numbers and its associated energy can be obtained as follow:
\begin{equation}
  \hat E_\mathrm{enc} = \sum_{\forall i} n_i \cdot e_i, 
  \label{eq:E_gen}
\end{equation}
where the index $i$ denotes the bit stream feature index, $n_i$ feature number, and $e_i$ the feature-specific encoding energy when one such feature occurs in decoding bit stream. In the case of the feature-based decoding energy estimator, the $n_i$ denotes the number of times a sub-process $i$ is performed, and $e_i$ refers to the energy consumed by the corresponding sub-process $i$. However, for encoding energy estimation using the bit stream features, $n_i$ feature number denotes the number of times a sub-process is performed while decoding, and the $e_i$ refers to the encoding energy consumed by the corresponding sub-process when that sub-process occurs once in decoding. 


The feature number derivation is explained in \cite{Herglotz18Journal}, and \cite{Herglotz16b}, where the feature counters implemented at positions corresponding to the standard's subclauses \cite{HEVCRec} were used. Using the syntax elements and variables used in the subclause \cite{HEVCRec}, \cite{Herglotz16b} defines a condition that must be maintained to increment the corresponding feature number and, in the end, implement the counters into any HEVC-conform decoder solution. \cite{Herglotz16b} introduces such a counter tool called HM Analyzer \cite{HManalyzer} based on the HEVC-Test Model (HM) reference software \cite{HM}. The HM Analyzer \cite{HManalyzer} yields very extensive bit stream features. As the complete set of bit stream features is rather elaborate and complex, we propose two models. Notably, we use a different set of features than the one used in \cite{Herglotz16b}, such that the features considered makes much sense for encoding. In addition, we select the features so that the difference between the measured and estimated energy is minimum. The first model is the elaborate model (EM), similar to the accurate model in \cite{Herglotz16b} which considers 100 bit stream features. The second one is a simple model (SM) that considers a reduced set of 50 features. Table \ref{tab:features} presents the features that are used in both models along with information on the features label, feature id, depth information, feature ID including depth information, and the tick on the columns SM and EM denotes the features using in SM and EM models respectively.

\begin{table}[]
\caption{Bit stream features used to estimate the energy consumption of a video encoder.The tick in the columns \textit{EM} and \textit{SM} shows if the feature is used in the elaborate and simple feature based model, respectively. }
\label{tab:features}
\centering
\begin{tabular}{|l|c|c|c|c|c|}
\hline
Feature label  & Depth (d) &ID incl. depth & \textit{SM} & \textit{EM} \\
\hline
$E_\mathrm{0}$ &- & 1& $\surd $ &$\surd $ \\
 Islice  &- & 2&$\surd $ & $\surd $ \\
 PBslice   &- & 3& $\surd $ &$\surd $  \\
\hline
 intraCU   &- & 4& $\surd$ & $\surd$  \\
 pla(d)   &4 & 5..8& - & $\surd $ \\
dc(d)   &4 & 9..12& - & $\surd $  \\
 hvd(d) &4 & 13..16   & - & $\surd $ \\
 ang(d) &4 & 17..20 & - & $\surd $ \\
 all(d)  &4 & 21..24  &$\surd $ & -  \\
 noMPM  &- & 25 & $\surd$ & $\surd$ \\
\hline
 skip(d) &4 & 26..29  & $\surd$ & $\surd$ \\
 merge(d) &4 & 30..33 & $\surd$ & $\surd$\\
 mergeSMP(d) &4 & 34..37  & $\surd$ & $\surd$\\
 mergeAMP(d) &4 & 38..39 & $\surd$ & $\surd$\\
 inter(d) &4 & 40..43& $\surd$ & $\surd$ \\
 interSMP(d)&4 & 44..47 & - & $\surd$ \\
 interAMP(d) &4 & 48..51& - & $\surd$\\
interCU(d)  &4 & 52..55 & $\surd$ & -\\
 fracpelHor(d) &4 & 56..59& - & $\surd$ \\
 fracpelVer(d) &4 & 60..63  & - & $\surd$  \\
 fracpelAvg  &- & 64& $\surd$ & -\\
 chrHalfpel(d) &4 & 65..68& $\surd$  & $\surd$\\
 bi  &- & 69& $\surd$ & $\surd$ \\
 MVD   &- & 70& $\surd$ & $\surd$ \\
 uni  &- & 71 & $\surd$ & $\surd$ \\
 fracopsHor  &- & 72 & - & $\surd$ \\
fracopsVer  &- & 73 & - & $\surd$ \\
fracopsBoth(d) &4 & 74..77  & $\surd$ & - \\
\hline
 coeff &- & 78& $\surd$ & $\surd$\\
 coeffg1 &- & 79 & - & $\surd$ \\
 CSBF  &- & 80 & - & $\surd$ \\
 val  &- & 81& $\surd$ & $\surd$ \\
 TrIntraY(d) &4 & 82..85  & - & $\surd$ \\
 TrIntraC(d) &4 & 85..88  & - & $\surd$ \\
 TrInterY(d) &4 & 89..92  & - & $\surd$\\
 TrInterC(d) &4 & 93..96  & - & $\surd$ \\
 Tr(d) &4 & 97..100 & $\surd$ & - \\
 TSF &- & 101& - & $\surd$\\
 zeroCoeff&- & 102 & - & $\surd$\\
\hline
 Bs0 &- & 103 & - & $\surd$ \\
 Bs1 &- & 104  & - & $\surd$ \\
 Bs2 &- & 105 & - & $\surd$ \\
 Bs  &- & 106& $\surd$ & - \\
 SAO\_Y\_BO  &- & 107& - & $\surd$\\
 SAO\_Y\_EO &- & 108&  - & $\surd$\\
 SAO\_Y  &- & 109& $\surd$ & - \\
 SAO\_C\_BO  &- & 110& - & $\surd$ \\
 SAO\_C\_EO &- & 111 & - & $\surd$ \\
 SAO\_C &- & 112&   $\surd$ & - \\
 SAO\_allComps &- & 113& - & $\surd$ \\
\hline
\end{tabular}
\end{table}

\section{Evaluation}\label{sec:eval}
\subsection{Encoding Energy Measurement Setup}\label{subsec:setup}
In this work, the energy demand of the encoding process is determined by two consecutive measurements, as explained in \cite{Herglotz18c}. First, the total energy consumed during the encoding process is measured. Then, the energy consumed in the idle state over the same duration of encoding is measured later. At last, the encoding energy $E_{\mathrm {enc}}$ is the difference between these two measurements.
\begin{equation} E_{\mathrm {enc}} =\int _{t=0}^{T} P_{\mathrm {total}}(t)dt- \int _{t=0}^{T} P_{\mathrm {idle}}(t)dt,  \end{equation}
where $T$ is the duration of the encoding process, $P_{\mathrm {total}}(t)$ is the total power consumed while encoding, $P_{\mathrm {idle}}(t)$ is the power consumed by the device in idle mode, and \textit{t} is the time. The confidence interval test for $m$ measurements with a standard deviation of $\sigma$ is defined as follows:
\begin{equation} \label{sta_test}2\cdot \frac {\sigma }{\sqrt {m}}\cdot t_\alpha (m-1) < \beta \cdot  E_{\mathrm{enc}}, \end{equation} 
where $\beta$ is the maximum encoding energy deviation, $\alpha$ is the probability with which \eqref{sta_test} is fulfilled and $t_\alpha$ denotes the critical t-value of Students's t-distribution. We chose $\alpha=0.99$ and $\beta=0.02$ and we repeat the measurements until the condition \eqref{sta_test} is satisfied. Hence, we ensure that $ E_{\mathrm{enc}}$ has a maximum energy deviation of 2\% from the actual energy consumed.
 \vspace{-0.1cm}
\subsection{Evaluation Sequences and Encoding Configurations}\label{subsec:sequences}
In this work, we perform multi-core encoding with the x265 encoder implementation \cite{x265}. We consider 22 sequences from the JVET common test conditions \cite{CTC} with various sequence characteristics such as frame rate, resolution, and content. Table \ref{tab:sequences} lists the sequences used in this work \cite{CTC}. In addition, we encode the first 64 frames of the sequences at different x265 presets, which are $\textit{ultrafast}$, $\textit{superfast}$, $\textit{veryfast}$, $\textit{faster}$, $\textit{fast}$, $\textit{medium}$, $\textit{slow}$, $\textit{slower}$, $\textit{veryslow}$, and various Constant Rate Factor (CRF) values, 18, 23, 28, 33. We generated 792 bit streams and further used them to train and validate the model. 


\begin{table}[]
    \centering
    \caption{Summary of sequences with its properties}
    \label{tab:sequences}
    \vspace{-0.15cm}
    \begin{tabular}{|c|c|c|c|}
    \hline
         \textbf{Sequence Name} & \textbf{Class} & \textbf{Resolution} & \textbf{Frame rate}\\
         \hline \hline
         PeopleOnStreet & A & 2560x1600 &30\\
         Traffic & A & 2560x1600 &30\\
         BasketballDrive & B & 1920x1080  & 50\\
         
         BQTerrace & B & 1920x1080  & 60\\
         
         Cactus & B & 1920x1080  & 50\\
         
         Kimono1 & B & 1920x1080  & 24\\
         
         ParkScene & B & 1920x1080  & 24\\
         
         BasketballDrill & C & 832x480  & 50\\
         
         BQMall & C & 832x480  & 60\\
         
         PartyScene & C & 832x480  & 50\\
         
         RaceHorses & C & 832x480  & 30\\
         
         BasketballPass & D & 416x240  & 50\\
         
         BlowingBubbles & D & 416x240  & 50\\
         
         BQSquare & D & 416x240  & 60\\
         
         RaceHorses & D & 416x240  & 30\\
         
         FourPeople & E & 1280x720  & 60\\
         
         Johnny & E & 1280x720  & 60\\
         
         KristenAndSara & E & 1280x720  & 60\\
            
         BasketballDrillText & F & 832x480 & 50\\
         
         ChinaSpeed & F & 1024x768 & 30\\
         
         SlideEditing & F & 1280x720 & 30\\
         
         Slideshow & F & 1280x720 & 20\\
         
         \hline
    \end{tabular}
\end{table}
 \vspace{-0.1cm}
\subsection{Model Evaluation}\label{subsec:eval}
We use mean relative estimation error for evaluation. By doing so, we get more significant results than using the absolute error as we strive to estimate the encoding energy accurately independent of the absolute energy, which can vary by several orders of magnitude. Thus, we show the relative estimation error of the measured encoding energy with respect to the estimated encoding energy for a single bit stream $n$ i.e., each bit stream $n$ corresponds to a single input sequence coded at a specific CRF, and for each preset $X$ as:
\begin{equation} \epsilon_{X,n}=\frac{\hat E_{\mathrm {enc}}- E_{\mathrm {enc}}}{ E_{\mathrm {enc}}} \Bigg\vert _{X,n} \end{equation}
where $\hat E_{\mathrm {enc}}$ is the estimated and $ E_{\mathrm{enc}}$ the measured encoding energy from (6). Then, we calculate the mean estimation error for each preset $X$ over each bit stream $n$ to obtain the overall estimation error for each preset:
\begin{equation} \overline{\epsilon}_{X}=\frac{1}{n}\sum_{n=1}^{n}\vert \epsilon_{X,n}\vert. \end{equation}

In order to determine the model parameter values for each preset, we perform a least-squares fit using a trust-region-reflective algorithm as presented in \cite{Coleman96}. We initially use the measured energies for a subset of the sequences referred to as the training set and their corresponding variables as input, which are the bit stream features. As a result,  we obtain the least-squares optimal parameters for the input training set, where we train the parameters such that the mean relative error is minimized. Ultimately, these model parameters are used to validate the model's accuracy on the remaining validation sequences. The training and validation data set are determined using a ten-fold cross-validation as proposed in \cite{Zaki14}. Using this technique, we randomly divide the complete set of measured energies into ten approximately equal-sized subsets. Then, for each subset, we use the other nine subsets to train the model, and the trained parameters are then used to validate the remaining subset by calculating the relative estimation error for all the sequences.
 \vspace{-0.1cm}
\subsection{Results and Discussion}\label{subsec:results}
\begin{figure}[]
\centering
 \input{res_eeenergy_cactus.tex}
 \includegraphics[width=0.5\textwidth]{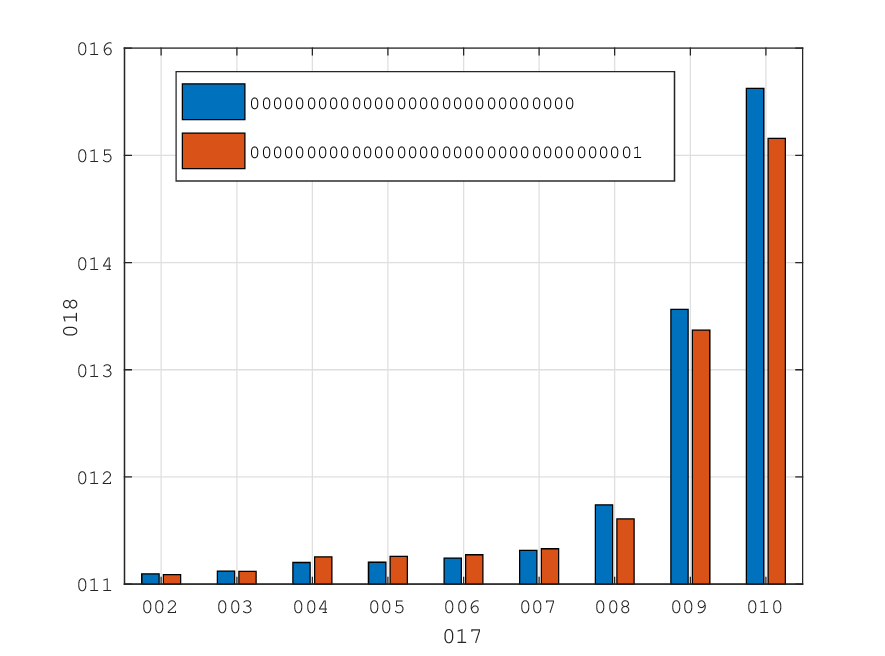}
 \vspace{-0.15cm}
  \caption{The measured energy $E_\mathrm{enc}$ and  the estimated energy $\hat E_\mathrm{enc}$ from SM for various presets 0-\textit{ultrafast}, 1-\textit{superfast}, 2-\textit{veryfast}, 3-\textit{faster}, 4-\textit{fast}, 5-\textit{medium}. 6-\textit{slow}, 7-\textit{slower}, 8-\textit{veryslow} for the sequence "Cactus" of the Class B.}
 \label{fig:superfast}
\end{figure}

The estimation errors of models for QP based model \cite{Rodriguez15}, the encoding time-based model \cite{Ramasubbu22} and the lightweight encoding time-based model \cite{Ramasubbu22}, along with the mean estimation errors for the bit stream feature-based models EM, and SM, are summarized in Table \ref{tab:ee}. The results show that the average estimation errors of models EM and SM perform best, with average errors smaller than 10\% for most presets. From the last two columns of table \ref{tab:ee}, we can see that, when all the presets are considered, i.e., when the training and validation is done over bit streams of all presets, unlike other models, both the EM and SM perform better. Notably, when considering each preset, the SM with 50 features performs better than the EM. The reason is that the EM results in overfitting, as seen in Table \ref{tab:ee} suggesting that in the case of preset known, we do not need the EM, and SM is enough for the better estimates. However, when considering all presets, overfitting is reduced to a great extent such that the EM outperforms the SM. In conclusion, we can say that EM should be used when the encoding preset is unknown, and SM can be used when the preset is known. For example, figure \ref{fig:superfast} shows the measured energy and estimation energy of all the presets from the SM for a Class B sequence "Cactus. "  The left bars correspond to the measured energy $E_\mathrm{enc}$, right bars to the estimated energy $\hat E_\mathrm{enc}$. Also, this figure shows the impact of various x265 presets on the encoding energy.
\begin{table}[]
\begin{small}
\caption{Relative mean estimation error for the QP based model(QP) \cite{Rodriguez15}, encoding time based model (T), lightweight (\textit{ultrafast}) encoding time model (UF) \cite{Ramasubbu22}, and proposed bistream based models EM and SM for all the bit streams. The lowest Relative mean estimation error across the different criteria is highlighted.}
\label{tab:ee}
\begin{tabular}{|l|r|r|r|r|r|}
\hline 
\textbf{Presets} & \textbf{QP} & \textbf{T}  & \textbf{UF}  & \textbf{EM}  & \textbf{SM}  \\ 
\hline 
\small \textit{ultrafast} & \small 32.82\% &\small 4.35\%& \small 4.35\% & 4.69\% & \textbf{3.49\%} \\ 
\small \textit{superfast} &\small 36.26\%  &\small \textbf{2.10\%}& \small 7.29\% & 4.84\% & 4.42\% \\ 
\small \textit{veryfast} &\small 38.99\% &\small 7.26\%&\small 9.61\% & 6.78\% & \textbf{4.06\%} \\ 
\small \textit{faster} &\small 36.73\%  & \small 7.37\% &\small 9.85\% & 7.05\% & \textbf{4.29\%} \\ 
\small \textit{fast} &\small 29.77\%  &\small \textbf{3.62\%}& \small 12.30\% & 7.80\% & 4.34\% \\ 
\small \textit{medium} &\small 37.71\%&\small 8.95\% & \small 10.96\% & 7.59\% & \textbf{5.65\% }\\ 
\small \textit{slow} & \small 33.53\%  &\small 7.40\%& \small 12.26\% & 12.22\% & \textbf{4.48\%} \\ 
\small \textit{slower}& \small 31.81\% & \small 10.71\% &\small 12.33\% & 7.78\% & \textbf{6.71\%} \\ 
\small \textit{veryslow} &\small 38.73\% & \small 11.58\%&\small 13.65\% & 9.27\% &\textbf{ 7.42\% }\\ 
\hline 
\textbf{average} &35.15\% &7.04\% &10.29\% &7.56\%&\textbf{4.88\%}\\
\hline
\textbf{all presets} &78.44\% & 15.86\%& 54.15\% & \textbf{13.56\%} & 14.77\% \\ 
\hline 
\end{tabular}
\end{small}
\end{table}
\section{Conclusion}\label{sec:concl}
Studying the energy demand of encoders is crucial for energy-efficient algorithms, but the energy measurements are time-consuming. Therefore, we need valid and simple energy estimators to overcome the drawback of such laborious measurements. With this respect, in this paper, we have shown that a bit stream feature-based approach to modeling the energy consumption of the HEVC software encoding process is highly suitable for estimating the encoding energy accurately. Moreover, the proposed model estimates the encoding energy with a mean estimation error of 4.88\% (averaged over presets), which outperforms models from the literature. The proposed model, coupled with the bit stream feature-based model for decoding, is useful in analyzing the energy efficiency of HEVC-coded bit streams, which is useful for identifying energy-efficient configurations. In future work, for an extended set of bit streams, we plan to count the sub-processes in encoding and obtain features-specific energies, address the energy demand of different categories of encoding sub-processes, and obtain the energy distribution of the encoding process.
\vspace{-0.15cm}
\bibliographystyle{IEEEtran}
\bibliography{IEEEabrv,literature.bib}
\end{document}

%% file: res_eeenergy_cactus.tex
%
\providecommand\matlabfragNegTickNoWidth{\makebox[0pt][r]{\ensuremath{-}}}%
%
%
\providecommand\matlabtextA{\color[rgb]{0.000,0.000,0.000}\fontsize{9.00}{9.00}\selectfont\strut}%
\psfrag{00000000000000000000000000000}[cl][cl]{\matlabtextA Measured energy, $E_{enc}$}%
\psfrag{00000000000000000000000000000000001}[cl][cl]{\matlabtextA Estimated energy, $\hat E_{enc}$}%
\providecommand\matlabtextB{\color[rgb]{0.150,0.150,0.150}\fontsize{11.00}{11.00}\selectfont\strut}%
\psfrag{017}[tc][tc]{\matlabtextB x265 presets}%
\psfrag{018}[bc][bc]{\matlabtextB Encoding energy in kJ}%
%
%
%
\providecommand\matlabtextC{\color[rgb]{0.150,0.150,0.150}\fontsize{10.00}{10.00}\selectfont\strut}%
\psfrag{002}[ct][ct]{\matlabtextC $0$}%
\psfrag{003}[ct][ct]{\matlabtextC $1$}%
\psfrag{004}[ct][ct]{\matlabtextC $2$}%
\psfrag{005}[ct][ct]{\matlabtextC $3$}%
\psfrag{006}[ct][ct]{\matlabtextC $4$}%
\psfrag{007}[ct][ct]{\matlabtextC $5$}%
\psfrag{008}[ct][ct]{\matlabtextC $6$}%
\psfrag{009}[ct][ct]{\matlabtextC $7$}%
\psfrag{010}[ct][ct]{\matlabtextC $8$}%
%
%
%
\psfrag{011}[rc][rc]{\matlabtextC $0$}%
\psfrag{012}[rc][rc]{\matlabtextC $5$}%
\psfrag{013}[rc][rc]{\matlabtextC $10$}%
\psfrag{014}[rc][rc]{\matlabtextC $15$}%
\psfrag{015}[rc][rc]{\matlabtextC $20$}%
\psfrag{016}[rc][rc]{\matlabtextC $25$}%
%